\documentclass[aip,twocolumn,apl,reprint]{revtex4-1}
\usepackage{graphicx}
\usepackage{amsmath}
\usepackage[T1]{fontenc}
\usepackage{dcolumn}
\usepackage{bm}
\usepackage[section]{placeins}

\newcommand{\Eq}[1]{Eq.(\ref{#1})}
\newcommand{\Fig}[1]{Fig.\,\ref{#1}}

\newcommand{\be}{\begin{equation}}
\newcommand{\ee}{\end{equation}}
\newcommand{\bd}{\begin{displaymath}}
\newcommand{\ed}{\end{displaymath}}

\newcommand{\siex}{\ensuremath{\sigma_{\rm ext}}}
\newcommand{\siexm}{\ensuremath{\bar{\sigma}_{\rm ext}}}
\newcommand{\siexs}{\ensuremath{\hat{\sigma}_{\rm total}}}
\newcommand{\siexsb}{\ensuremath{\hat{\sigma}_{\rm noise}}}
\newcommand{\siexsr}{\ensuremath{\hat{\sigma}_{\rm ext}}}

\newcommand{\sisc}{\ensuremath{\sigma_{\rm sca}}}
\newcommand{\sia}{\ensuremath{\sigma_{\rm abs}}}

\newcommand{\sic}{\ensuremath{\sigma_{\rm c}}}

\newcommand{\NA}{\ensuremath{{\rm NA}}}
\newcommand{\NAc}{\ensuremath{{\rm NA_c}}}
\newcommand{\dc}{\ensuremath{d_{\rm c}}}
\newcommand{\If}{\ensuremath{I_{\rm f}}}
\newcommand{\Id}{\ensuremath{I_{\rm d}}}
\newcommand{\Ai}{\ensuremath{A_{\rm i}}}
\newcommand{\Ab}{\ensuremath{A_{\rm b}}}
\newcommand{\Db}{\ensuremath{\Delta_{\rm b}}}
\newcommand{\ri}{\ensuremath{r_{\rm i}}}
\newcommand{\rd}{\ensuremath{r_{\rm d}}}

\newcommand{\Is}{\ensuremath{I_{\rm df}}}

\newcommand{\Nfw}{\ensuremath{N_{\rm fw}}}
\newcommand{\Na}{\ensuremath{N_{\rm a}}}

\newcommand{\dpix}{\ensuremath{d_{\rm px}}}
\newcommand{\Npix}{\ensuremath{N_{\rm px}}}
\newcommand{\nph}{\ensuremath{N_{\rm ph}}}

\newcommand{\nm}{\rm nm}
\newcommand{\ext}{\rm ext}

\begin{document}
\title{Polarization-resolved extinction and scattering cross-section of individual gold
nanoparticles measured by wide-field microscopy on a large ensemble}
\author{Lukas M Payne}
\affiliation{School of Biosciences, Cardiff University, Cardiff CF24
3AA, United Kingdom}

\author{Wolfgang Langbein}
\affiliation{School of Physics and Astronomy, Cardiff University, the Parade, Cardiff
CF24 3AA, United Kingdom}

\author{Paola Borri}
\email[Electronic address:]{BorriP@cardiff.ac.uk}
\affiliation{School of Biosciences, Cardiff University, Cardiff CF24
3AA, United Kingdom}

\date{\today}

\begin{abstract}

We report a simple, rapid, and quantitative wide-field technique to measure the optical extinction
\siex\ and scattering \sisc\ cross-section  of single nanoparticles using wide-field microscopy
enabling simultaneous acquisition of hundreds of nanoparticles for statistical analysis. As a
proof of principle, we measured nominally spherical gold nanoparticles of 40\,nm and 100\,nm
diameter and found mean values and standard deviations of \siex\ and \sisc\ consistent with
previous literature. Switching from unpolarized to linearly polarized excitation, we measured
\siex\ as a function of the polarization direction,  and used it to characterize the asphericity
of the nanoparticles. The method can be implemented cost-effectively on any conventional
wide-field microscope and is applicable to any nanoparticles.

\end{abstract}
 \maketitle

Metallic nanoparticles (NPs) exhibit morphology-dependent localized surface plasmon resonances
(LSPR) which couple to propagating light and manifest as an increased particle polarisability at
the LSPR frequency. Besides fundamental interest, these local optical resonances can be exploited
to image metallic NPs with high sensitivity and to probe nanoscale regions in the NP vicinity via
the local field enhancement effect, with possible applications ranging from sub-wavelength optical
devices\,\cite{BarnesN03}, catalysis\cite{AwazuJACS08} and photovoltaics\,\cite{AtwaterNM10} to
biomedical imaging\,\cite{CognetAC08,MasiaOL09} and
sensing\,\cite{KneippJPCM02,MacFarlandNL03,SoennichsenNB05}.

Important physical quantities characterizing the linear optical properties of a NP are the
absorption cross-section \sia, the scattering cross-section \sisc, and the resulting extinction
cross-section $\siex=\sia+\sisc$. Beyond traditional ensemble average measurements, a number of
approaches have been developed recently to measure \siex\ at the single particle level, showing
that the optical properties of individual NPs can significantly differ from the ensemble average
owing to inhomogeneities in NP size and shape. It is therefore particularly important to develop a
technique able to rapidly quantify the cross-sections at the single NP level and to perform a
statistical analysis over many NPs, providing a relevant sample characterization. Quantitative
values of \siex, \sia, and/or \sisc\ of single NPs have been reported using dark-field
micro-spectroscopy\,\cite{AndersonJPC10}, photothermal imaging\,\cite{TcherniakNL10}, and spatial
modulation micro-spectroscopy\,\cite{MuskensPRB08}. However, in order to provide cross-section
values in absolute units, dark-field micro-spectroscopy and photothermal imaging require a
calibration reference (e.g. by comparison with \siex known from theory) while spatial modulation
micro-spectroscopy needs a precise measurement of the beam profile at the sample. Moreover,
photothermal imaging and spatial modulation micro-spectroscopy are beam-scanning techniques,
therefore costly and less amenable to the rapid characterization of a large number of NPs compared
to wide-field techniques. Additionally they are modulation-based which requires specialized
equipment such as acousto-optical modulators and lock-in detection.

In this work, we report a simple and quantitative wide-field technique to measure \siex\ and
\sisc\ on single NPs using conventional bright and dark-field microscopy without the need of
calibration standards and with a field of view enabling simultaneous acquisition of hundreds of
NPs.

The experimental set-up consists of an inverted microscope (Nikon Ti-U) equipped with a
white-light illumination (halogen lamp 100W with Nikon neutral color balance filter), a oil
condenser of 1.4 numerical aperture (\NA) with a removable home-built dark-field illumination of
1.1-1.4\,NA, a 40x 0.95\,NA dry objective, a 1.5x intermediate magnification, and a Canon EOS 40D
color camera attached to the left port of the microscope. Images were taken in Canon 14-bit RAW
format with 10.1 megapixel resolution. The investigated samples were gold NPs (GNPs) of nominal
40\,nm and 100\,nm diameter (BBInternational) covalently bound onto a glass coverslip
functionalised with (3-mercapto) triethoxysilane, covered in silicone oil (refractive index ${\rm
n}=1.518$) and sealed with a glass slide.

Dark-field microscopy was performed initially to confirm the
presence of metallic NPs appearing as colored scatterers,
distinguishable from the white scattering of other debris or glass
roughness. The color camera enables a coarse spectroscopic detection
separating the three wavelength ranges \cite{Ratledge12} of red (R)
570-650\,nm, green (G) 480-580\,nm, and blue (B) 420-510\,nm. Once a
suitable region was located and focussed, a dark-field image was
taken. Subsequent bright-field microscopy was performed by removing
the dark field ring and adjusting the condenser numerical aperture
\NAc\ to match the objective \NA. To quantitatively measure the
extinction cross-sections, two bright-field transmission images were
taken, one with the NPs in the objective focus and the second one
out-of-focus, moving the objective by approximately $d=15\,\mu$m
axially away from the sample. Background images were taken for
blocked illumination. To achieve the lowest shot noise, the lowest
camera sensitivity was used (100 ISO), for which the full-well
capacity of the pixels of about $\Nfw=4\times 10^4$ electrons occurs
at 70\% of the digitizer range (3.4 electrons/count). The exposure
time in the order of 10\,ms was chosen to reach $\Nfw$. Averaging
over $\Na=36$ acquisitions was performed for each image set.

\begin{figure}
\includegraphics[width=\columnwidth]{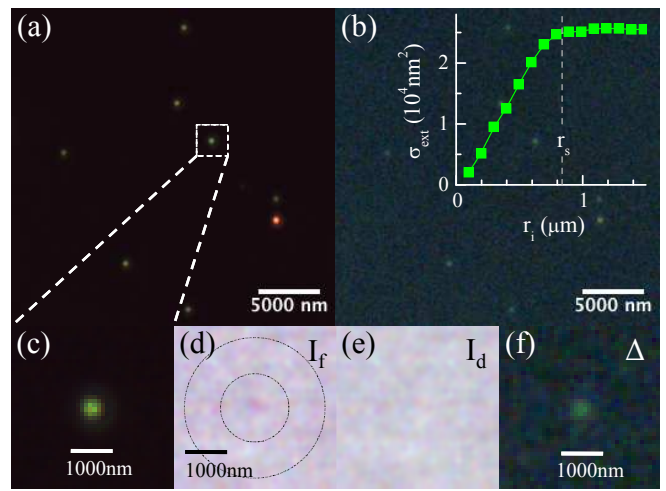}
\caption{(a) Full Color (FC) dark-field image of 40\,nm diameter
gold NPs. (b) Corresponding FC extinction image $\Delta$ from 0
(black) to 0.043 (white). (c) Zoom of dark-field image. (d,e)
Corresponding zoom of the FC bright-field transmission with NP in
focus \If\ (d) and out-of-focus \Id\ (e). (f) Zoom of FC extinction
image.}\label{fig:NP_dark_ext}
\end{figure}

Let us call the background-subtracted transmitted intensity of the bright-field image with NPs in
focus \If\, and the defocused intensity \Id. In the defocused image a NP distributes its effect
over a radius of about $\rd=\NA d$ making \Id\ similar to the intensity \If\, in the absence of
the NP. The extinction cross-section of a NP located within the area \Ai\ in the image can then be
expressed as $\siex =\int_{\Ai}\Delta dA$ with the relative extinction $\Delta=(\Id-\If)/\Id$. An
example of a full color dark-field image and the corresponding $\Delta$ image for gold NPs of
40\,nm diameter is shown in \Fig{fig:NP_dark_ext}. To account for the slight mismatch between \Id\
and $\If$ without NP due to the defocusing and the residual influence of the NP, we determine a
local background extinction $\Db=\Ab^{-1}\int_{\Ab}\Delta dA$ in the area \Ab\ between the radius
$\ri$ and $2\ri$, as sketched in \Fig{fig:NP_dark_ext}d, yielding the background-corrected $\siex
=\int_{\Ai}(\Delta-\Db) dA$. The correction area \Ab\ is within the defocused image of the NP,
i.e. $2\ri<\rd$, ensuring a homogeneous influence of the NP over \Ai\ and \Ab. The dependence of
the measured \siex\ on \ri\ is shown in the inset of \Fig{fig:NP_dark_ext}b for the G channel,
using a constant \Db\, from $\ri=1.5\,\mu$m . A saturation of \siex\ is observed for $\ri>800\,\nm
\sim 3\lambda/(2\NA)$, approximately at the second Airy ring of the objective point-spread
function. This behavior can be qualitatively understood considering that $\Delta$ is the result of
the interference between the scattered field of the NP and the illumination field. For a spatially
coherent illumination this interference would lead to fringes in \siex\ decaying as $1/\sqrt{r}$.
However, the short spatial coherence length $\dc\sim\lambda/\NAc$ of the illumination is
suppressing these fringes for $r>\dc$.

In order to determine \siex\ of many particles from an extinction image we developed an image
analysis programme written in ImageJ macro language. We split the raw images into RGB channels,
subtract the background, average the multiple acquisitions of \If\ and \Id, and calculate
$\Delta$. We then determine the particle locations as the maxima of $\Delta$ with values in a
range adjusted to reject noise and large aggregates. For each maximum we choose \Ai\ given by a
centered disk of radius $\ri=3\lambda/(2\NA)=837$\,nm and calculate \siex.

\begin{figure}
\includegraphics[width=\columnwidth]{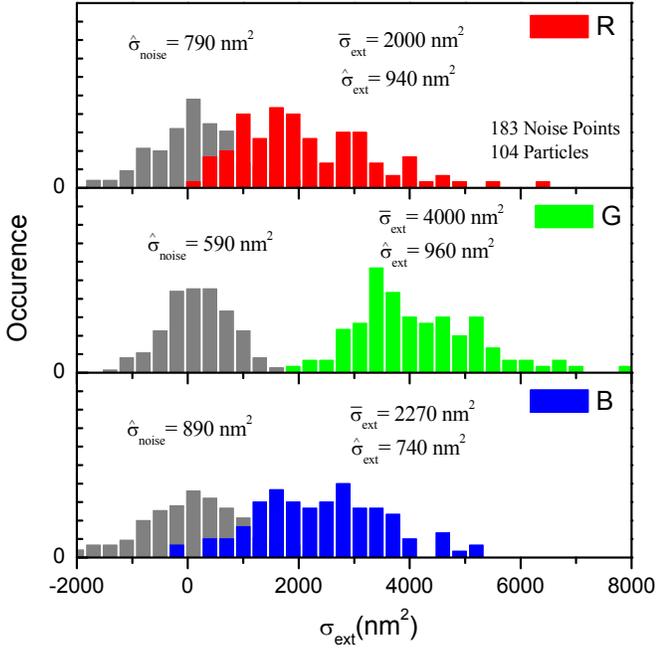}
\caption{Distributions of $\siex$ for 104 GNPs of nominal 40\,nm diameter in the R, G, and B color
channels as indicated. The grey histograms show the noise distribution obtained by measuring
$\siex$ in 183 randomly selected regions without NPs.}\label{fig:Hist_40}
\end{figure}

We select individual NPs using their extinction color, retaining NPs with \siex\ largest
in the color channel corresponding to the expected plasmon resonance, i.e. green (red)
for 40\,nm (100\,nm) spherical GNPs having a LSPR at 540\,nm (590\,nm) in a surrounding
medium of 1.5 refractive index. NPs which likely correspond to aggregates, debris, or
largely non-spherical GNPs are excluded in this way. The statistical results over 104
individual GNPs of nominally 40\,nm diameter are summarised in \Fig{fig:Hist_40}. The
distribution of \siex\ in the G channel corresponding to the plasmon resonance has a mean
of $\siexm=4000\,\nm^2$, which is consistent with experimental and theoretical values
found in literature\cite{VanDijkPCCP06,NoguezOM05,VanDijkPhD07,MuskensPRB08}. The
standard deviation $\siexs=1300\,\nm^2$ of \siex\ contains a part \siexsb\ due to
measurement noise. This part is determined using the distribution of \siex\ in image
regions not containing NPs, resulting in a zero mean and a standard deviation \siexsb.
The standard deviation \siexsr\ arising from the NPs is accordingly determined by
$\siexsr^2=\siexs^2-\siexsb^2$. We find $\siexsb=590\,\nm^2$ and $\siexsr=960\,\nm^2$ for
the G channel. $\siexsr$ can be attributed to a size distribution of the GNPs as follows.
The scaling of $\siex \propto R^\gamma$ for spherical particles of radius $R$ is known
from Mie theory\,\cite{MieADP08}. In the dipole approximation, $\gamma \approx 3$ for
small particles where the extinction is dominated by absorption and increases towards
$\gamma=6$ for larger particles where the extinction is dominated by scattering. We found
$\gamma \approx 3$  for 40\,nm diameter at 532\,nm wavelength using calculated absorption
and scattering cross-sections \cite{VanDijkPCCP06}. This scaling allows us to estimate
the relative standard deviation of the radius $\delta R/R =\siexsr/(\siexm \gamma)
\approx 0.08$. The manufacturer specifies $\delta R/R<0.08$ for 40\,nm and 100\,nm
particles determined by electron microscopy. Thus $\siexsr$ is on the upper limit of what
expected from the size distribution of spherical particles in a constant dielectric
environment. It has been shown in the literature that additional factors influencing
$\siexsr$ might be the NP non perfect sphericity\,\cite{TcherniakNL10}, as well as
fluctuations in the local dielectric environment and the electron-surface scattering
damping parameter\,\cite{MuskensPRB08}. Measurements of $\siex$ for 100\,nm GNPs (not
shown) in the red channel yield $\siexm=30000\,\nm^2$ and $\siexsr=4558\,\nm^2$,
resulting in $\delta R/R=0.034$ with $\gamma=4.5$. These values are consistent with
literature \cite{VanDijkPCCP06,NoguezOM05,VanDijkPhD07,MuskensPRB08} for spherical
100\,nm GNPs and meet the manufacturers size specifications.

\begin{figure}
\includegraphics[width=0.95\columnwidth]{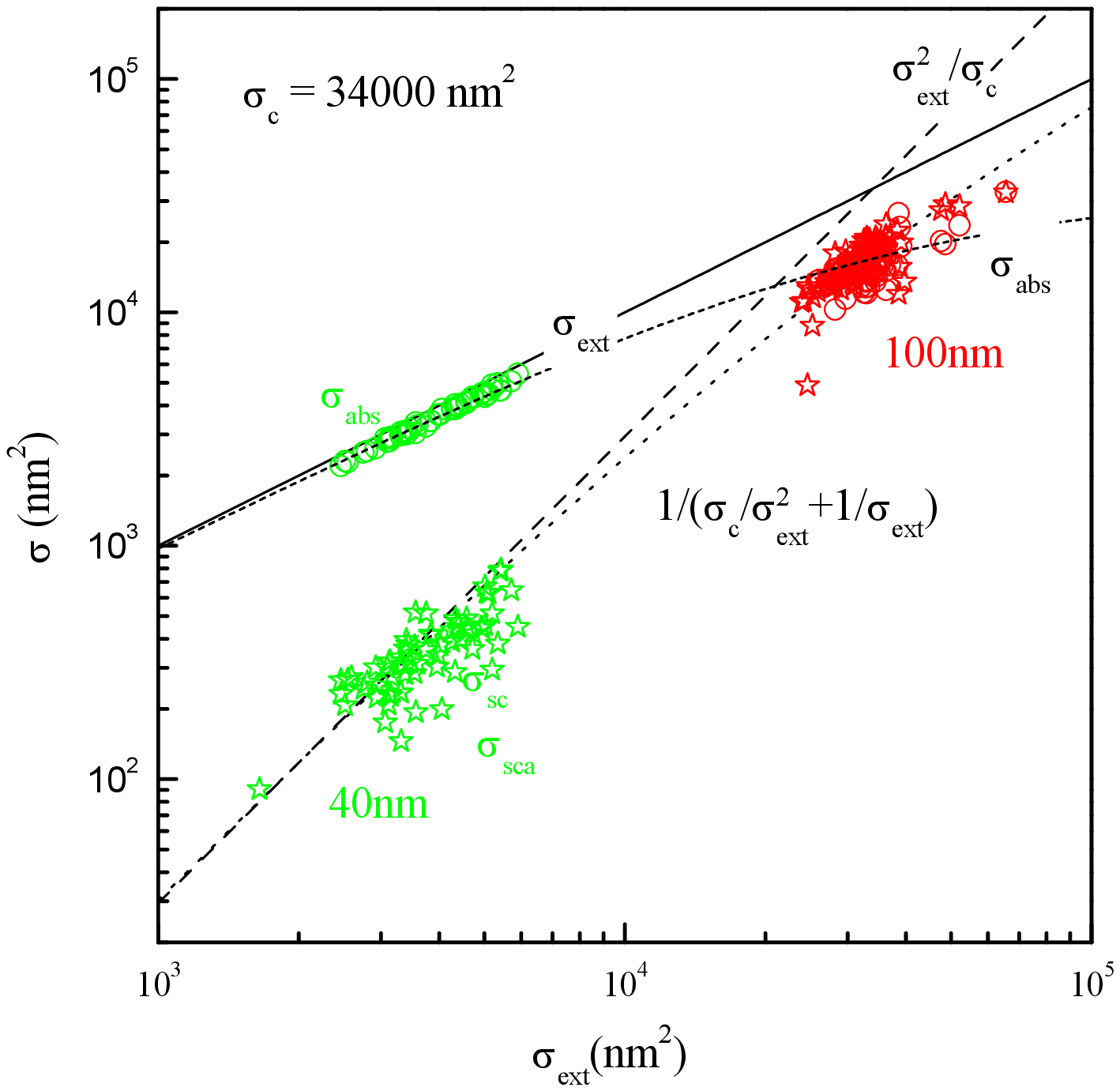}
\caption{Scattering cross-section $\sisc$ (stars) measured in dark-field images versus
$\siex$ measured in extinction images for 40\,nm GNPs and 100\,nm GNPs. The deduced
absorption crossections \sia\ are given as circles. The dotted line gives the fitted
scaling as labeled. The limiting behaviour for small $\siex$ (dashed line) and large
$\siex$ (solid line), and the expected absorption cross-section (short-dashed) are also
shown.}\label{fig:NP_scat}
\end{figure}

Using the scattered intensity \Is\ measured in dark-field microscopy integrated over the
same spatial area \Ai\ as \siex, we obtain the detected power scattered by the NP which
is proportional to the scattering cross-section \sisc. Normalizing the scattered
intensity to \Id\, we can write $\sisc=\eta\int_{\Ai} \Is dA/\Id$ with the constant
$\eta$ determined only by the condenser \NAc\ ranges in bright field and dark field and
the objective \NA. If $\eta$ is known \sisc\ can be quantified in absolute units. We
determined $\eta$ as follows. We compared \sisc\ with \siex\ on each NP of the ensemble,
as shown in \Fig{fig:NP_scat}. GNPs with \siex\ well below a certain cut-off value (\sic)
are dominated by absorption with cross-section \sia\ scaling like $R^3$. Since \sisc\ is
proportional to $R^6$ in this regime, we expect $\sisc\sic = \siex^2$. Conversely,
scattering dominates for larger particles such that $\siex\simeq \sisc$. This trend is
indeed observed in \Fig{fig:NP_scat}, and allows us to infer $\sic=34000\,\nm^2$ and in
turn determine $\eta=26$, both with about 10\% error. The resulting \sisc\ of the GNPs is
consistent with literature\cite{VanDijkPhD07}. Furthermore, we can deduce the absorption
cross-section $\sia=\siex-\sisc$, which is also shown in \Fig{fig:NP_scat}. Using the
calibrated \sisc\, we find a detection limit for \sisc\ of about $100\,\nm^2$ due to
background scattering contributing to \Is\ in our samples. The camera dark noise allows
in principle to detect $\sisc<1\,\nm^2$.

While the detection limit for \sisc\ is given by the sample background scattering, the
detection limit for \siex\ is given by the shot noise in the measured transmitted
intensity.  The relative shot noise is given by $1/\sqrt{\nph}$ with the detected number
of photons \nph\, which is determined by the number of acquisitions $\Na$, the full-well
capacity $\Nfw$ of the camera pixels, the number of pixels $\Npix$ in the area $\Ai$, and
the fraction $\nu$ of pixels used for the color channel (for the Bayer color filter of
our camera $\nu=1/2$ for G and $\nu=1/4$ for R, B), yielding
$\siexsb=\Ai/\sqrt{\Na\Nfw\nu\Npix}$. With the pixel size $\dpix$, the area
$\Ai=\pi\ri^2$ with $\ri=3\lambda/(2\NA)$, and the magnification $M$ onto the camera, we
find
\be \label{eq:noise} \siexsb=\frac{3\lambda\dpix}{2 M\NA}\sqrt{\frac{\pi}{\Na\Nfw\nu}}\ee
For the green channel of \Fig{fig:Hist_40}, we have $\Nfw=4\times10^4$, $\Na=36$, $M=60$,
$\dpix=5.7\,\mu$m, $\NA=0.95$, $\lambda=0.53\,\mu$m, and $\nu=1/2$, yielding $\siexsb=589\,\nm^2$,
in agreement with the measured noise. The blue and red channels have a factor of $\sqrt{2}$ larger
noise due to the smaller $\nu$. This detection limit could be improved by using an oil immersion
objective with 1.45\,\NA, $M=150$, and $\Na=1800$ possible in a 60 second video, for which
\Eq{eq:noise} yields $\siexsb=43\,\nm^2$, which would allow measuring single GNPs down to 10\,nm
diameter.

We note that the finite angular range of the objective implies that it collects also a fraction of
the scattered light leading to an underestimate of the extinction. The solid angle in the samples
with 1.5 refractive index collected by the objective is 1.6\,sterad, which for isotropic
scattering is 13\% of the scattered light. We could correct for this by adding 13\% of the
measured \sisc\ to \siex. We also note that \sisc\ is determined using the scattering of the
dark-field excitation into the objective, which also has a certain angular range that needs to be
considered if the scattering is sufficiently anisotropic.

\begin{figure}
\includegraphics[width=\columnwidth]{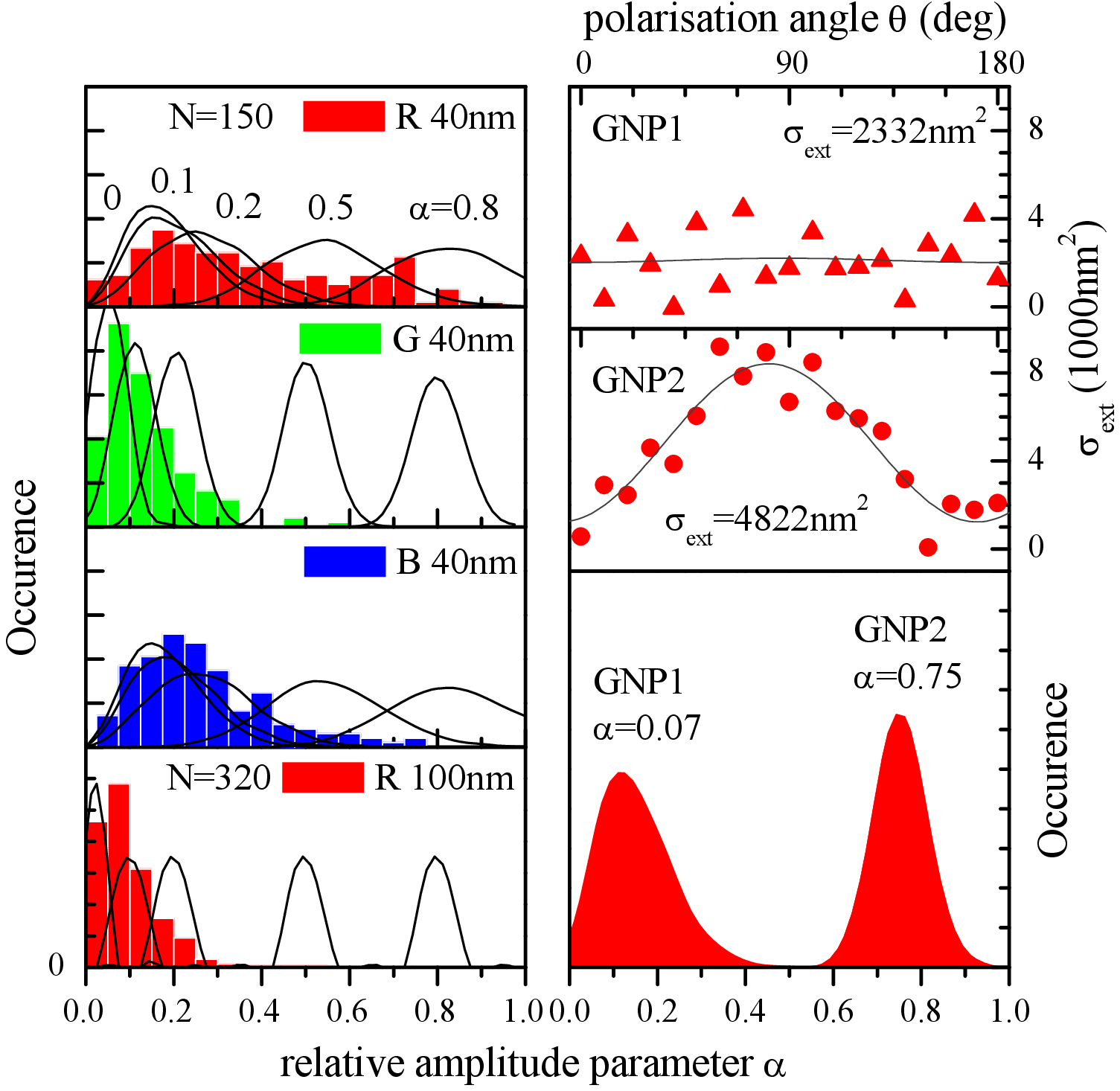}
\caption{Left Column: Distributions of the amplitude $\alpha$, for 150 GNPs of 40\,nm diameter in
the R, G, and B color channel, and 320 GNPs of 100\,nm diameter in the R channel. The
distributions due to \siexsb\ for GNPs of $\alpha=0,0.1,0.2,0.5,0.8$ are given as black lines
using \siexm\ as given in \Fig{fig:Hist_40}. Right Column: $\siex(\theta)$ in the R channel with
fits for two 40\,nm GNPs, and distribution of the deduced amplitude paramater $\alpha$ by the
measurement noise \siexsb.}\label{fig:NP_pol}
\end{figure}

Furthermore, we measured the dependence of $\siex$ on the linear polarization angle $\theta$ of
the excitation light, which is a sensitive probe of NP asphericity, by inserting a linear
polariser in the illumination beam path before the condenser. The resulting $\siex(\theta)$  is
shown in \Fig{fig:NP_pol}(right) for $\theta$ between $0^{\circ}$ to $180^{\circ}$ in steps of
$10^{\circ}$ for two selected GNPs in the red channel. We analyze these results by fitting the
expression $\sigma_{\ext}(\theta)=\sigma_0(1+\alpha\,\cos(2(\theta-\theta_0)))$ where $\sigma_0$
is the polarization-averaged $\siex$, $\alpha \ge 0$ is the amplitude of the polarization
dependence, and $0\ge \theta_0 \ge \pi$ is an angular offset, indicating the direction of the NP
asymmetry. To estimate the influence of the measurement noise on the fit parameters we calculated
their distribution over Gaussian random fluctuations of the fitted $\siex(\theta)$ with a standard
deviation $\siexsb$. GNP1 has a fitted $\alpha=0.07$, and its distribution has a mean value
$\bar{\alpha}=0.15$ and a standard deviation $\hat{\alpha}=0.08$. GNP2 instead is significantly
non-spherical with a fitted $\alpha=0.75$ and a distribution with $\bar{\alpha}=0.75$ well above
$\hat{\alpha}=0.06$. The red channel is used here as it is most sensitive to LSPR shifts due to
asphericity.

The distribution of $\alpha$ over the NP ensemble is shown in \Fig{fig:NP_pol} (left) for 40\,nm
and 100\,nm GNPs for different color channels. For comparison, the simulated distribution of
$\alpha$ for GNPs having $\siex(\theta)$ given by the fit function are shown for
$\alpha=0,0.1,0.2,0.5,0.8$ in \Fig{fig:NP_pol} as black lines using $\sigma_0=\siexm$ of the color
channel. The comparison shows that the polarization dependence can identify non-spherical GNPs
through the distinct values of $\alpha$. To further infer the NP geometrical aspect ratio from
these data a comparison with theory is needed which will be reported in a forthcoming work.

In conclusion, we have shown that conventional wide-field microscopy can be implemented with a
consumer camera to extract quantitative values of polarization-resolved extinction, scattering and
absorption cross-sections of individual nanoparticles and generate histograms for statistical
characterization of large numbers of particles. Although quasi-spherical gold nanoparticles of
40\,nm and 100\,nm diameter were used in this work for proof of principle, the technique is
applicable to any nanoparticles, with a detection sensitivity limit in the order of $100\,\nm^2$.
Importantly, this technique can be adopted by any laboratory equipped with conventional wide-field
microscopy as a tool to quantify the linear optical response of a statistically significant number
of individual nanoparticles.

This work was supported by the UK EPSRC Research Council under the Leadership fellowship award of
P.B., grant n. EP/I005072/1 and EP/I016260/1.

\bibliography{ExtMethod,langsrv}

\begin{thebibliography}{16}%
\makeatletter
\providecommand \@ifxundefined [1]{%
 \@ifx{#1\undefined}
}%
\providecommand \@ifnum [1]{%
 \ifnum #1\expandafter \@firstoftwo
 \else \expandafter \@secondoftwo
 \fi
}%
\providecommand \@ifx [1]{%
 \ifx #1\expandafter \@firstoftwo
 \else \expandafter \@secondoftwo
 \fi
}%
\providecommand \natexlab [1]{#1}%
\providecommand \enquote  [1]{``#1''}%
\providecommand \bibnamefont  [1]{#1}%
\providecommand \bibfnamefont [1]{#1}%
\providecommand \citenamefont [1]{#1}%
\providecommand \href@noop [0]{\@secondoftwo}%
\providecommand \href [0]{\begingroup \@sanitize@url \@href}%
\providecommand \@href[1]{\@@startlink{#1}\@@href}%
\providecommand \@@href[1]{\endgroup#1\@@endlink}%
\providecommand \@sanitize@url [0]{\catcode `\\12\catcode `\$12\catcode
  `\&12\catcode `\#12\catcode `\^12\catcode `\_12\catcode `\%12\relax}%
\providecommand \@@startlink[1]{}%
\providecommand \@@endlink[0]{}%
\providecommand \url  [0]{\begingroup\@sanitize@url \@url }%
\providecommand \@url [1]{\endgroup\@href {#1}{\urlprefix }}%
\providecommand \urlprefix  [0]{URL }%
\providecommand \Eprint [0]{\href }%
\providecommand \doibase [0]{http://dx.doi.org/}%
\providecommand \selectlanguage [0]{\@gobble}%
\providecommand \bibinfo  [0]{\@secondoftwo}%
\providecommand \bibfield  [0]{\@secondoftwo}%
\providecommand \translation [1]{[#1]}%
\providecommand \BibitemOpen [0]{}%
\providecommand \bibitemStop [0]{}%
\providecommand \bibitemNoStop [0]{.\EOS\space}%
\providecommand \EOS [0]{\spacefactor3000\relax}%
\providecommand \BibitemShut  [1]{\csname bibitem#1\endcsname}%
\let\auto@bib@innerbib\@empty
\bibitem [{\citenamefont {Barnes}, \citenamefont {Deraux},\ and\ \citenamefont
  {Ebbesen}(2003)}]{BarnesN03}%
  \BibitemOpen
  \bibfield  {author} {\bibinfo {author} {\bibfnamefont {W.}~\bibnamefont
  {Barnes}}, \bibinfo {author} {\bibfnamefont {A.}~\bibnamefont {Deraux}}, \
  and\ \bibinfo {author} {\bibfnamefont {T.}~\bibnamefont {Ebbesen}},\
  }\href@noop {} {\bibfield  {journal} {\bibinfo  {journal} {Nature}\ }\textbf
  {\bibinfo {volume} {424}},\ \bibinfo {pages} {824} (\bibinfo {year}
  {2003})}\BibitemShut {NoStop}%
\bibitem [{\citenamefont {Awazu}\ \emph {et~al.}(2008)\citenamefont {Awazu},
  \citenamefont {Fujimaki}, \citenamefont {Rockstuhl}, \citenamefont
  {Tominaga}, \citenamefont {Murakami}, \citenamefont {Ohki}, \citenamefont
  {Yoshida},\ and\ \citenamefont {Watanabe}}]{AwazuJACS08}%
  \BibitemOpen
  \bibfield  {author} {\bibinfo {author} {\bibfnamefont {K.}~\bibnamefont
  {Awazu}}, \bibinfo {author} {\bibfnamefont {M.}~\bibnamefont {Fujimaki}},
  \bibinfo {author} {\bibfnamefont {C.}~\bibnamefont {Rockstuhl}}, \bibinfo
  {author} {\bibfnamefont {J.}~\bibnamefont {Tominaga}}, \bibinfo {author}
  {\bibfnamefont {H.}~\bibnamefont {Murakami}}, \bibinfo {author}
  {\bibfnamefont {Y.}~\bibnamefont {Ohki}}, \bibinfo {author} {\bibfnamefont
  {N.}~\bibnamefont {Yoshida}}, \ and\ \bibinfo {author} {\bibfnamefont
  {T.}~\bibnamefont {Watanabe}},\ }\href@noop {} {\bibfield  {journal}
  {\bibinfo  {journal} {J. Am. Chem. Soc.}\ }\textbf {\bibinfo {volume}
  {130}},\ \bibinfo {pages} {1676} (\bibinfo {year} {2008})}\BibitemShut
  {NoStop}%
\bibitem [{\citenamefont {Atwater}\ and\ \citenamefont
  {Polman}(2010)}]{AtwaterNM10}%
  \BibitemOpen
  \bibfield  {author} {\bibinfo {author} {\bibfnamefont {H.}~\bibnamefont
  {Atwater}}\ and\ \bibinfo {author} {\bibfnamefont {A.}~\bibnamefont
  {Polman}},\ }\href@noop {} {\bibfield  {journal} {\bibinfo  {journal} {Nat.
  Mater.}\ }\textbf {\bibinfo {volume} {9}},\ \bibinfo {pages} {205} (\bibinfo
  {year} {2010})}\BibitemShut {NoStop}%
\bibitem [{\citenamefont {Cognet}\ \emph {et~al.}(2008)\citenamefont {Cognet},
  \citenamefont {Berciaud}, \citenamefont {Lasne},\ and\ \citenamefont
  {Lounis}}]{CognetAC08}%
  \BibitemOpen
  \bibfield  {author} {\bibinfo {author} {\bibfnamefont {L.}~\bibnamefont
  {Cognet}}, \bibinfo {author} {\bibfnamefont {S.}~\bibnamefont {Berciaud}},
  \bibinfo {author} {\bibfnamefont {D.}~\bibnamefont {Lasne}}, \ and\ \bibinfo
  {author} {\bibfnamefont {B.}~\bibnamefont {Lounis}},\ }\href@noop {}
  {\bibfield  {journal} {\bibinfo  {journal} {Anal. Chem.}\ }\textbf {\bibinfo
  {volume} {80}},\ \bibinfo {pages} {2288} (\bibinfo {year}
  {2008})}\BibitemShut {NoStop}%
\bibitem [{\citenamefont {Masia}\ \emph {et~al.}(2009)\citenamefont {Masia},
  \citenamefont {Langbein}, \citenamefont {Watson},\ and\ \citenamefont
  {Borri}}]{MasiaOL09}%
  \BibitemOpen
  \bibfield  {author} {\bibinfo {author} {\bibfnamefont {F.}~\bibnamefont
  {Masia}}, \bibinfo {author} {\bibfnamefont {W.}~\bibnamefont {Langbein}},
  \bibinfo {author} {\bibfnamefont {P.}~\bibnamefont {Watson}}, \ and\ \bibinfo
  {author} {\bibfnamefont {P.}~\bibnamefont {Borri}},\ }\href@noop {}
  {\bibfield  {journal} {\bibinfo  {journal} {Opt. Lett.}\ }\textbf {\bibinfo
  {volume} {34}},\ \bibinfo {pages} {1816} (\bibinfo {year}
  {2009})}\BibitemShut {NoStop}%
\bibitem [{\citenamefont {Kneipp}\ \emph {et~al.}(2002)\citenamefont {Kneipp},
  \citenamefont {Kneipp}, \citenamefont {Itzkan}, \citenamefont {Dasari},\ and\
  \citenamefont {Feld}}]{KneippJPCM02}%
  \BibitemOpen
  \bibfield  {author} {\bibinfo {author} {\bibfnamefont {K.}~\bibnamefont
  {Kneipp}}, \bibinfo {author} {\bibfnamefont {H.}~\bibnamefont {Kneipp}},
  \bibinfo {author} {\bibfnamefont {I.}~\bibnamefont {Itzkan}}, \bibinfo
  {author} {\bibfnamefont {R.}~\bibnamefont {Dasari}}, \ and\ \bibinfo {author}
  {\bibfnamefont {M.}~\bibnamefont {Feld}},\ }\href@noop {} {\bibfield
  {journal} {\bibinfo  {journal} {J. Phys.: Condens. Matter}\ }\textbf
  {\bibinfo {volume} {14}},\ \bibinfo {pages} {R597} (\bibinfo {year}
  {2002})}\BibitemShut {NoStop}%
\bibitem [{\citenamefont {McFarland}\ and\ \citenamefont
  {Van~Duyne}(2003)}]{MacFarlandNL03}%
  \BibitemOpen
  \bibfield  {author} {\bibinfo {author} {\bibfnamefont {A.}~\bibnamefont
  {McFarland}}\ and\ \bibinfo {author} {\bibfnamefont {R.}~\bibnamefont
  {Van~Duyne}},\ }\href@noop {} {\bibfield  {journal} {\bibinfo  {journal}
  {Nano Lett.}\ }\textbf {\bibinfo {volume} {3}},\ \bibinfo {pages} {1057}
  (\bibinfo {year} {2003})}\BibitemShut {NoStop}%
\bibitem [{\citenamefont {Sonnichsen}\ \emph {et~al.}(2005)\citenamefont
  {Sonnichsen}, \citenamefont {Reinhard}, \citenamefont {Liphardt},\ and\
  \citenamefont {Alivisatos}}]{SoennichsenNB05}%
  \BibitemOpen
  \bibfield  {author} {\bibinfo {author} {\bibfnamefont {C.}~\bibnamefont
  {Sonnichsen}}, \bibinfo {author} {\bibfnamefont {B.}~\bibnamefont
  {Reinhard}}, \bibinfo {author} {\bibfnamefont {J.}~\bibnamefont {Liphardt}},
  \ and\ \bibinfo {author} {\bibfnamefont {A.}~\bibnamefont {Alivisatos}},\
  }\href@noop {} {\bibfield  {journal} {\bibinfo  {journal} {Nature Biotech.}\
  }\textbf {\bibinfo {volume} {23}},\ \bibinfo {pages} {741} (\bibinfo {year}
  {2005})}\BibitemShut {NoStop}%
\bibitem [{\citenamefont {Anderson}\ \emph {et~al.}(2010)\citenamefont
  {Anderson}, \citenamefont {Mayer}, \citenamefont {Fraleigh}, \citenamefont
  {Yang}, \citenamefont {Lee},\ and\ \citenamefont {Hafner}}]{AndersonJPC10}%
  \BibitemOpen
  \bibfield  {author} {\bibinfo {author} {\bibfnamefont {L.}~\bibnamefont
  {Anderson}}, \bibinfo {author} {\bibfnamefont {K.}~\bibnamefont {Mayer}},
  \bibinfo {author} {\bibfnamefont {R.}~\bibnamefont {Fraleigh}}, \bibinfo
  {author} {\bibfnamefont {Y.}~\bibnamefont {Yang}}, \bibinfo {author}
  {\bibfnamefont {S.}~\bibnamefont {Lee}}, \ and\ \bibinfo {author}
  {\bibfnamefont {J.}~\bibnamefont {Hafner}},\ }\href@noop {} {\bibfield
  {journal} {\bibinfo  {journal} {J. Phys. Chem. C}\ }\textbf {\bibinfo
  {volume} {114}},\ \bibinfo {pages} {11127} (\bibinfo {year}
  {2010})}\BibitemShut {NoStop}%
\bibitem [{\citenamefont {Tcherniak}\ \emph {et~al.}(2010)\citenamefont
  {Tcherniak}, \citenamefont {Ha}, \citenamefont {Dominguez-Medina},
  \citenamefont {Slaughter},\ and\ \citenamefont {Link}}]{TcherniakNL10}%
  \BibitemOpen
  \bibfield  {author} {\bibinfo {author} {\bibfnamefont {A.}~\bibnamefont
  {Tcherniak}}, \bibinfo {author} {\bibfnamefont {J.}~\bibnamefont {Ha}},
  \bibinfo {author} {\bibfnamefont {S.}~\bibnamefont {Dominguez-Medina}},
  \bibinfo {author} {\bibfnamefont {L.}~\bibnamefont {Slaughter}}, \ and\
  \bibinfo {author} {\bibfnamefont {S.}~\bibnamefont {Link}},\ }\href@noop {}
  {\bibfield  {journal} {\bibinfo  {journal} {Nano Lett.}\ }\textbf {\bibinfo
  {volume} {10}},\ \bibinfo {pages} {1398} (\bibinfo {year}
  {2010})}\BibitemShut {NoStop}%
\bibitem [{\citenamefont {Muskens}\ \emph {et~al.}(2008)\citenamefont
  {Muskens}, \citenamefont {Billaud}, \citenamefont {Broyer}, \citenamefont
  {Del~Fatti},\ and\ \citenamefont {Vallée}}]{MuskensPRB08}%
  \BibitemOpen
  \bibfield  {author} {\bibinfo {author} {\bibfnamefont {O.}~\bibnamefont
  {Muskens}}, \bibinfo {author} {\bibfnamefont {P.}~\bibnamefont {Billaud}},
  \bibinfo {author} {\bibfnamefont {M.}~\bibnamefont {Broyer}}, \bibinfo
  {author} {\bibfnamefont {N.}~\bibnamefont {Del~Fatti}}, \ and\ \bibinfo
  {author} {\bibfnamefont {F.}~\bibnamefont {Vallée}},\ }\href@noop {}
  {\bibfield  {journal} {\bibinfo  {journal} {Phys. Rev. B}\ }\textbf {\bibinfo
  {volume} {78}},\ \bibinfo {pages} {205410} (\bibinfo {year}
  {2008})}\BibitemShut {NoStop}%
\bibitem [{\citenamefont {Ratledge}(2012)}]{Ratledge12}%
  \BibitemOpen
  \bibfield  {author} {\bibinfo {author} {\bibfnamefont {D.}~\bibnamefont
  {Ratledge}},\ }\href {http://www.deep-sky.co.uk/imaging/dslr/dslr.htm}
  {\enquote {\bibinfo {title} {Digital slr imaging},}\ }\bibinfo {howpublished}
  {http://www.deep-sky.co.uk/imaging/dslr/dslr.htm} (\bibinfo {year}
  {2012})\BibitemShut {NoStop}%
\bibitem [{\citenamefont {van Dijk}\ \emph {et~al.}(2006)\citenamefont {van
  Dijk}, \citenamefont {Tchebotareva}, \citenamefont {Orrit}, \citenamefont
  {Lippitz}, \citenamefont {Berciaud}, \citenamefont {Lasne}, \citenamefont
  {Cognet},\ and\ \citenamefont {Lounis}}]{VanDijkPCCP06}%
  \BibitemOpen
  \bibfield  {author} {\bibinfo {author} {\bibfnamefont {M.}~\bibnamefont {van
  Dijk}}, \bibinfo {author} {\bibfnamefont {A.}~\bibnamefont {Tchebotareva}},
  \bibinfo {author} {\bibfnamefont {M.}~\bibnamefont {Orrit}}, \bibinfo
  {author} {\bibfnamefont {M.}~\bibnamefont {Lippitz}}, \bibinfo {author}
  {\bibfnamefont {S.}~\bibnamefont {Berciaud}}, \bibinfo {author}
  {\bibfnamefont {D.}~\bibnamefont {Lasne}}, \bibinfo {author} {\bibfnamefont
  {L.}~\bibnamefont {Cognet}}, \ and\ \bibinfo {author} {\bibfnamefont
  {B.}~\bibnamefont {Lounis}},\ }\href@noop {} {\bibfield  {journal} {\bibinfo
  {journal} {Phys. Chem. Chem. Phys.}\ }\textbf {\bibinfo {volume} {8}},\
  \bibinfo {pages} {2486} (\bibinfo {year} {2006})}\BibitemShut {NoStop}%
\bibitem [{\citenamefont {Noguez}(2005)}]{NoguezOM05}%
  \BibitemOpen
  \bibfield  {author} {\bibinfo {author} {\bibfnamefont {C.}~\bibnamefont
  {Noguez}},\ }\href@noop {} {\bibfield  {journal} {\bibinfo  {journal}
  {Optical Materials}\ }\textbf {\bibinfo {volume} {27}},\ \bibinfo {pages}
  {1204} (\bibinfo {year} {2005})}\BibitemShut {NoStop}%
\bibitem [{\citenamefont {van Dijk}(2007)}]{VanDijkPhD07}%
  \BibitemOpen
  \bibfield  {author} {\bibinfo {author} {\bibfnamefont {M.}~\bibnamefont {van
  Dijk}},\ }\emph {\bibinfo {title} {Nonlinear-optical studies of single gold
  nanoparticles}},\ \href@noop {} {Ph.D. thesis},\ \bibinfo  {school}
  {Universiteit Leiden} (\bibinfo {year} {2007})\BibitemShut {NoStop}%
\bibitem [{\citenamefont {Mie}(1908)}]{MieADP08}%
  \BibitemOpen
  \bibfield  {author} {\bibinfo {author} {\bibfnamefont {G.}~\bibnamefont
  {Mie}},\ }\href@noop {} {\bibfield  {journal} {\bibinfo  {journal} {Annalen
  der Physik}\ }\textbf {\bibinfo {volume} {330}},\ \bibinfo {pages} {377}
  (\bibinfo {year} {1908})}\BibitemShut {NoStop}%
\end{thebibliography}%

\end{document}